\documentclass[english]{revtex4}
\usepackage[T1]{fontenc}
\usepackage[latin1]{inputenc}

\makeatletter

\usepackage{babel}
\makeatother
\begin{document}

\title{Oskar Klein, the sixth dimension and the strength of a magnetic pole}

\author{{\small A. Sandoval-Villalbazo}$^{a}${\small , L.S. García-Colín}$^{b,\, c}${\small and
A.L. García-Perciante}$^{a}$}

\address{$^{a}${\small Departamento de Física y Matemáticas, Universidad
Iberoamericana, México D.F.}}

\address{$^{b}${\small Departamento de Física, Universidad Aut\'{o}noma
Metropolitana, México D.F.}}

\address{$^{c}${\small El Colegio Nacional, Centro Histórico, M\'{e}xico
D.F.}}

\begin{abstract}
This work extends to six dimensions the idea first proposed by Klein
regarding a closed space in the context of a fifth dimension and its
link to quantum theory. The main result is a formula that expresses
the value of the characteristic length of the sixth dimension in terms
of the strength of a magnetic monopole $g$. It is shown that in the
case of Dirac's monopole, the ratio of the characteristic lengths
of the fifth and sixth dimension corresponds to twice the fine structure
constant $\alpha$. Possible consequences of the idea are discussed.
\end{abstract}
\maketitle

\section{Introduction}

It is well known that the world lines of charged particles in the
presence of gravitational and electromagnetic fields can be viewed
as geodesics in the 5D space-time first proposed by T. Kaluza \cite{Kaluza1}.
Kaluza's treatment was later improved by Klein in 1926 \cite{Klein1},
in a quantum theoretical approach, giving rise to the so-called Kaluza-Klein
theory \cite{Klein2}. This formalism also makes possible to propose,
based on plausible grounds, a simple formula that gives the value
of the characteristic length of the fifth dimension in terms of fundamental
constants of nature. The original Kaluza-Klein theories have been,
since the 1920's, the subject of further analysis that include generalizations
to a larger number of dimensions \cite{Cremmer}, cosmological implications
\cite{Salam} and magnetic monopoles \cite{mono1}, most of them treated
\textit{a-la} Dirac \cite{Dirac2}. 

This last subject deserves more attention. Recently \cite{JMP1},
the authors have observed that a simple 6D generalization of Kaluza's
5D metric leads naturally to the geodesics of particles possessing
(hypothetical) fundamental magnetic charges. Further, this 6D space-time
reproduces Maxwell's equations in the presence of magnetic monopoles
and allows the establishment of a wave equation for the vector potential
\cite{Nos2}. In view of these facts, we have reanalyzed Klein's formalism
in order to establish a link between the eigenvalue of the sixth dimensional
momentum operator and the discrete nature of the hypothetical point
charges. We also examine the characteristic length of the sixth dimension
by setting the magnetic charge equal to the value proposed by Dirac.

\section{Kaluza's classic theory in 6 dimensions}

We start this section observing that the geodesic equation 

\begin{equation}
\frac{d^{2}x^{\alpha}}{dt^{2}}+\Gamma_{\beta\lambda}^{\alpha}\frac{dx^{\beta}}{dt}\frac{dx^{\lambda}}{dt}=0\label{uno}\end{equation}
 contains the equation of motion of a magnetic charged particle in
the presence of a \emph{static} electromagnetic field given that: 

\begin{equation}
x^{\alpha}=\left[\begin{array}{c}
x^{1}\\
x^{2}\\
x^{3}\\
ct\\
0\\
\frac{1}{c\xi}\frac{g}{m}t\end{array}\right]\label{dos}\end{equation}

\begin{equation}
\dot{x}^{\alpha}=\frac{dx^{\alpha}}{dt}\label{tres}\end{equation}

\begin{equation}
g_{\mu\nu}=\left[\begin{array}{cccccc}
1 & 0 & 0 & 0 & 0 & 0\\
0 & 1 & 0 & 0 & 0 & 0\\
0 & 0 & 1 & 0 & 0 & 0\\
0 & 0 & 0 & -1 & -\xi\varphi & -\frac{\xi}{c}\eta\\
0 & 0 & 0 & -\xi\varphi & 1 & 0\\
0 & 0 & 0 & -\frac{\xi}{c}\eta & 0 & 1\end{array}\right]\label{cuatro}\end{equation}
In Eq. (\ref{uno}), the indices $\alpha$, $\beta$ and $\gamma$
run from $1$ to $6$ and \[
\xi=\sqrt{\frac{16\pi G\epsilon_{0}}{c^{2}}},\]
is introduced in order to obtain the correct gravitational and electromagnetic
static limits (Poisson equations) as shown in the appendix of Ref.
\cite{GR}. Also, in Eqs. (\ref{dos}-\ref{cuatro}), $\varepsilon_{o}=8.85\times10^{-12}\frac{c^{2}}{N\, m^{2}},\, G=6.67\times10^{-11}\frac{N\, m^{2}}{kg^{2}}$
and $\, c=3\times10^{8}\frac{m}{s}$. Dimensional consistency requires
a magnetic charge $g$ measured in $\frac{m}{s}c$. The metric (\ref{cuatro})
involves the static potentials $\varphi$ and $\eta$, and can easily
be extended to include non-static fields \cite{Kaluza1}-\cite{JMP1}.
Nevertheless, for the ideas here presented, the static fields metric
will be enough.

\section{Klein's hypothesis applied to $x^{6}$}

Following Klein \cite{Klein2}, we write for the six dimensional momentum
the equation:

\begin{equation}
p^{6}=m_{g}\dot{x}^{6}=\frac{g}{c\xi}\label{cincoB}\end{equation}
If we now apply the standard quantization rules to the quantity defined
in Eq. (\ref{cincoB}), then:

\begin{equation}
\frac{g}{c\xi}=\frac{N_{g}h}{\ell_{6}}\label{seisB}\end{equation}
where $N_{g}$ is an integer number, as required by the quantization
of charge, and $\ell_{6}$ represents the characteristic length of
the (compact) periodic six dimension. Use has been made of Eqs. (\ref{dos}-\ref{tres}).
Equation (\ref{seisB}) is a natural extension of Klein's arguments
regarding the sixth dimension and expresses the idea that its characteristic
length is inversely proportional to the value of the magnetic monopole
strength $g$.

\section{Characterization of the sixth dimension}

There is no experimental evidence to fix the two parameters in Eq.
(\ref{seisB}): the characteristic length $\ell_{6}$ and the strength
of the magnetic charge $g$. In this sense we propose two alternatives.
Firstly, an assumption can be made about the size of the sixth dimension
compared to $\ell_{5}$ which is given by\begin{equation}
\ell_{5}=\frac{h}{e}\sqrt{\frac{16\pi G\varepsilon_{0}}{c^{2}}}\label{l5}\end{equation}
If both compact dimensions posses the same characteristic length,
then $g$ is simply given by $g=qc$. If otherwise, we assume a vanishing
value for $g$, then $\ell_{6}$ would become infinite. This idea
seems consistent with the existence of non-compact extra dimensions.
If compactness is a property sought for dimensional spaces of $d>4$,
the existence of the still undetected magnetic monopole would favour
the idea. 

The second alternative is to give a value for $g$ which fixes the
length $\ell_{6}$. Most work regarding Kaluza-Klein theory in the
context of magnetic monopoles is based on Dirac's pioneering work
\cite{Dirac2}. In that context, the magnetic monopole possesses a
charge $g=\frac{\varepsilon_{o}hc^{2}}{e}=3.3\times10^{-9}C\frac{m}{s}$
and thus,\begin{equation}
\ell_{6}=\frac{e}{\varepsilon_{o}c^{2}}\sqrt{16\pi G\varepsilon_{o}}\label{L6}\end{equation}
 In this case, by fixing the magnitude of the magnetic charge, one
can determine the size of the sixth dimension which is smaller than
the fifth dimension as determined by Klein. Moreover, the ratio between
the two extra dimensions is found to be:\begin{equation}
\frac{\ell_{6}}{\ell_{5}}=\frac{e^{2}}{h\varepsilon_{o}c}=2\alpha\label{siete}\end{equation}
 where $\alpha$ is the fine structure constant, which has been found
to be related with the length of the extra dimensions in the Kaluza-Klein
context by other authors \cite{freund}.

\section{Summary and discussion}

In previous work, we've shown that Kaluza's theory can be extended
to six dimensions in order to include an elementary magnetic charge
in electromagnetic theory. By doing so, Maxwell's equations become
symmetric and can be derived directly from Einstein's field equations.
Here, the theory formulated in Ref. \cite{JMP1} has been complemented
by applying the quantization hypothesis proposed by Klein for the
fifth dimension to the new sixth dimension. As a result, an expression
for the characteristic length of the sixth dimension is expressed
in terms of the strength of the magnetic charge {[}see Eq. (\ref{L6}){]}.
In order to estimate either of these quantities we propose two alternatives.
If $\ell_{6}$ is compared directly with $\ell_{5}$, a value for
the magnetic charge is obtained. On the other hand, if $g$ is assumed
to correspond to the value of Dirac's monopole, the size of the sixth
dimension that can be obtained is related with $\ell_{5}$ through
the fine structure constant. 

Most work extending Kaluza's ideas is based on Gauge theories and
topological arguments. The results here presented arise from a simpler
methodology in which the dimensionality of space-time is increased
in the same way Kaluza did in his classic paper. Hence, Klein's hypothesis
can be applied directly to the extra dimension and thus, the analysis
can be done in a straightforward and simple manner.

Kaluza's idea has been abandoned for years for various reasons, one
of them being the mass spectra of the particles obtained in five dimensions
\cite{Straumann}. In this sense we wish to point out that the topology
in 6 dimensions may differ with the one in which that calculation
is based. Calculating the mass spectra in a new topology is a task
worth pursuing since, as some drawbacks of Kaluza's classical theory
were surmounted by extending the metric in Ref. \cite{JMP1}, new
elements may enter the calculation for the mass spectra and modify
the results whereas more symmetry is achieved.

It is the opinion of the authors that physical ideas leading to simple
measurable properties of spaces of more than four dimensions must
be pursued. Part of this work has already been introduced in the form
of a Kaluza-Klein magnetohydrodynamical framework \cite{Nos3,GR}.
We believe that the present work is another step in this direction. 

This work has been supported by CONACyT (Mexico), project 41081-F
and FICSAC (Mexico), PFSA.

\end{document}